# Toy Blocks and Rotational Physics

**Gabriele U. Varieschi and Isabel R. Jully,** Loyola Marymount University, Los Angeles, CA

Have you ever observed a child playing with toy blocks? A favorite game is to build towers and then make them topple like falling trees. To the eye of a trained physicist this should immediately look like an example of the physics of "falling chimneys," when tall structures bend and break in mid-air while falling to the ground. The game played with toy blocks can actually reproduce well what is usually seen in photographs of falling towers, such as the one that appeared on the cover of the September 1976 issue of The Physics Teacher.[1] In this paper we describe how we performed and analyzed these simple but interesting experiments with toy blocks.

One of us recently published a detailed paper,[2] summarizing the physics of the rotational motion of tall structures, falling under gravity. In this analysis it is assumed that the structure falls while maintaining a point of contact at the bottom, so that its motion is essentially a rotation around this pivot point, and the only external forces are the weight **W** of the body and a constraint force **F** at the base (see Fig. 1). We performed experiments using models made with toy blocks, and filmed the falling motion with a digital video camera. These video clips and still pictures of the models can be viewed on a related web page.[3]

A classical physics demonstration, usually called the "falling stick" or "hinged stick," works on the same principle, showing that the acceleration of gravity can be exceeded during the fall by some points of the body (see again the review paper[2] and references therein). The difference between a falling stick and a real falling tower (or the toy models we used) is that the latter is not rigid and will usually bend and break during the fall. This can be seen in Fig. 2, and it is well documented by all the photos and videos on our web site.[3]

### The causes of the breaking

The internal forces causing the tower to bend and break are also shown in Fig. 1. We use polar coordinates r and θ, to describe the rotational motion of the body around the point of contact with the ground at the origin, for a structure of total height H. For an arbitrary cross section of the tower, at a distance r from the origin, the internal forces can be modeled as a longitudinal stress force **P** (either a tension or a compression), a transverse shearing force **S**, and a "bending moment" $\mathbf{N_b}$, represented by the curved arrow in Fig. 1. This bending moment is the main cause of the bending and breaking of the structure, and can be thought of as originating from a "pair" of forces applied at the leading and trailing edges of the tower.

Complete details about the calculation of these forces are given in Ref. 2. Summarizing these results, we observe that the tower can break in two possible ways. In the first case, a particularly strong transverse shear force **S** can "cut" the tower. This is more likely to happen for real structures (such as falling chimneys, towers, etc.) and the breaking typically occurs near the bottom, where **S** is larger. In the second case, by far the most common with small-scale models, the structure progressively bends and breaks, due to a combined effect of the bending moment **N_b** and the longitudinal force **P**.

This combined effect is usually better described by the longitudinal stress at the leading edge $\sigma_L$, which can be computed from the previous quantities, and depends on the tilt angle $\theta$, the height fraction r/H and the side a of the square cross section of the tower:[4]

$$\frac{\sigma_L a^2}{W} = -\frac{1}{2}\left(1 - \frac{r}{H}\right)\left[\left(5 + 3\frac{r}{H}\right)\cos\theta - 3\left(1 + \frac{r}{H}\right)\right] + \frac{3}{2}\frac{H}{a}\sin\theta\frac{r}{H}\left(1 - \frac{r}{H}\right)^2. \qquad (1)$$

In this equation the leading edge stress $\sigma_L$ is normalized, by dividing by the factor $W/a^2$, in order to obtain a dimensionless quantity. For a given value of the ratio H/a (total height H divided by the side a of the tower), Eq. (1) relates $\sigma_L$ directly to the variables $\theta$ and r/H. By plotting this function of two variables, it is easy to see which values of $\theta$ and r/H will maximize $\sigma_L$, determining the angle and height fraction at which the structure is more likely to break.

**Toy block models**

To check the theory outlined above, we set up models built with different types of toy blocks. We induced their fall, making sure they would start rotating around an appropriate support at the bottom. Details of the blocks we used and the dimensions of the models can be found in Refs. 2 and 3. These experiments are very simple and inexpensive and can be made part of a laboratory class devoted to rotational mechanics or used as physics demonstrations.

In order to improve on our previous work, we used video capture software (VideoPoint 2.5), together with a digital video camera, to record and analyze the motion. An example of a picture frame taken from one of these video recordings is shown in Fig. 2, while the complete video-clips can be viewed on our web page.[3]

In Fig. 2 we observe the fall of a tower made of 24 cubic wooden blocks (H/a = 24), which appears to bend around the ninth block from the bottom (for r/H ≅ 0.354). The use of the software allows for a better determination of the breaking angle $\theta$, which is estimated to be around 20°-25°. As can be seen from the picture, the software can follow the different rotations of the top and bottom portions of the structure, showing the point where the tower begins to bend.

In Fig. 3 we plot the quantities of Eq. 1, for the case of our toy model with H/a = 24. The normalized stress at the leading edge is shown as a function of the height ratio r/H, for different tilt angles. For a given angle, the structure should break at the point where the stress is maximum (corresponding to the solid points in the figure).

We notice a good agreement with the behavior of our toy model from Fig. 2. For a breaking angle around 20°-25°, Fig. 3 predicts a breaking height ratio r/H ≅ 0.35-0.37 (considering the maxima of the red curves and the related dashed green lines in Fig.3). Our structure seems in fact to be bending and breaking at r/H ≅ 0.354, as mentioned above.

Similar experiments can be performed with different types of blocks, or varying dimensions of the towers, and will typically show results consistent with the theory outlined here.

**Conclusion**

Toy blocks can easily be used to illustrate some peculiar effects of rotational dynamics. These experiments can be effectively integrated into classroom or lab activities at various levels. They can be shown as simple demonstrations of rotational physics in introductory classes, or become more challenging activities for advanced mechanics courses or laboratories.

**Acknowledgments**

This research was supported by an award from Research Corporation. The authors would like to thank Dr. V. Coletta for the useful discussions regarding this paper.

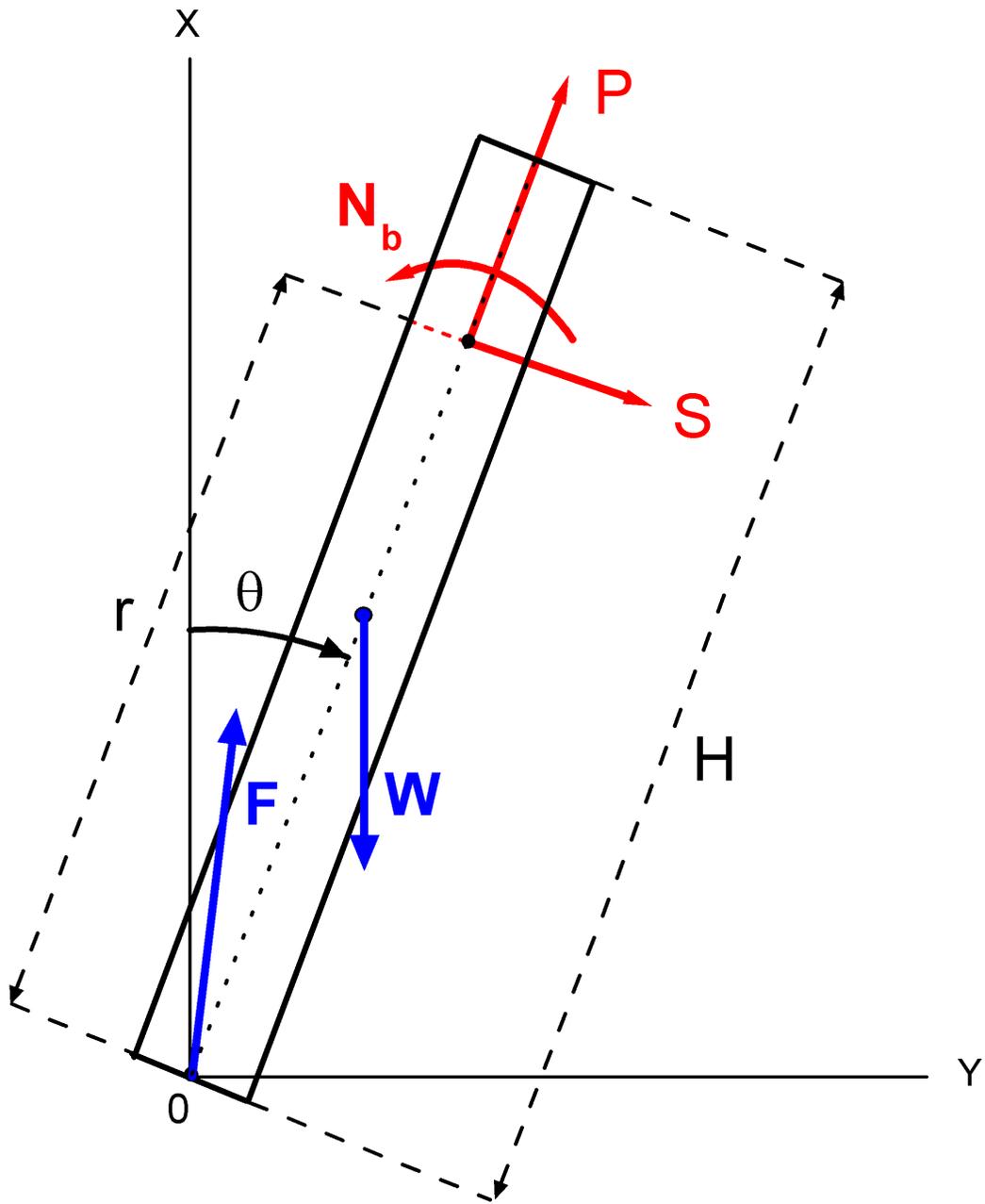

Fig. 1. The falling tower or chimney described as a rotating uniform stick. The external forces, the weight **W** and the constraint force **F** at the base, are shown in blue. The internal forces (**P**, **S**) and the bending moment **N$_b$**, at an arbitrary cross section, are shown in red.

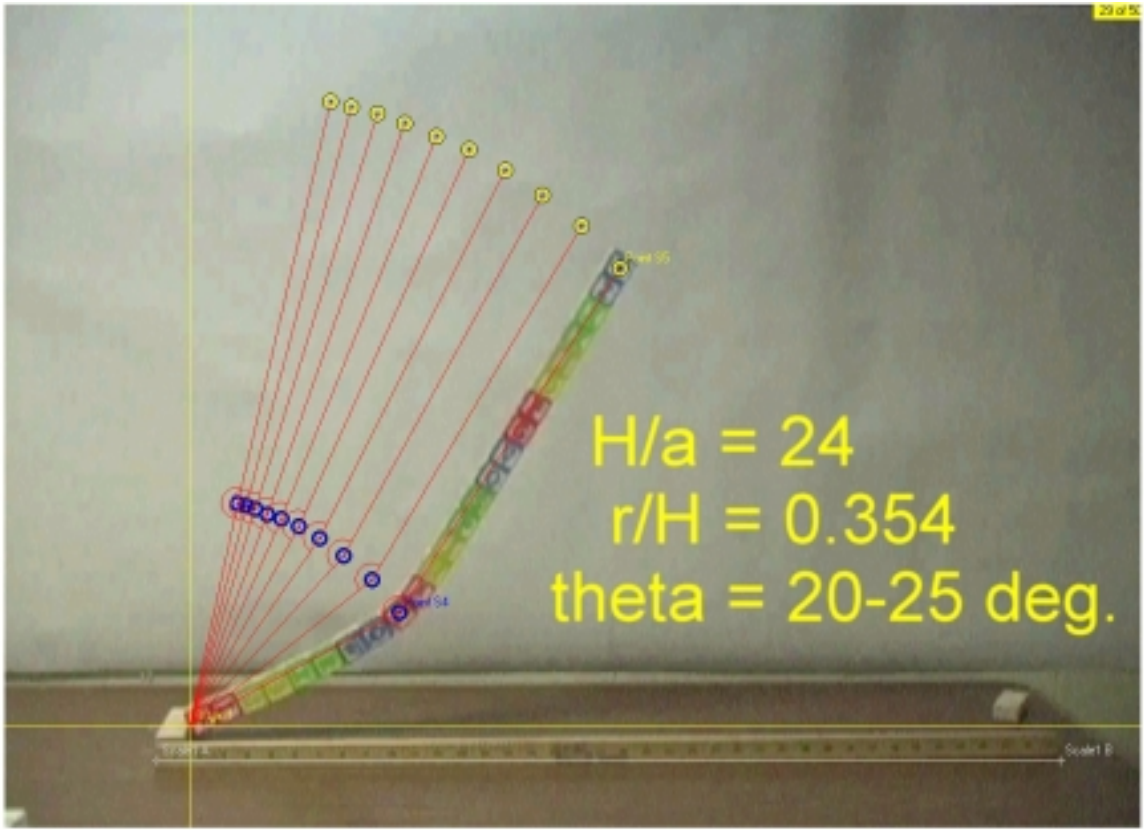

Fig. 2. A toy model made with wooden blocks. The structure appears to bend and break at r/H ≅ 0.354 and at an angle θ = 20°-25°. VideoPoint software is used to show the progressive bending of the tower.

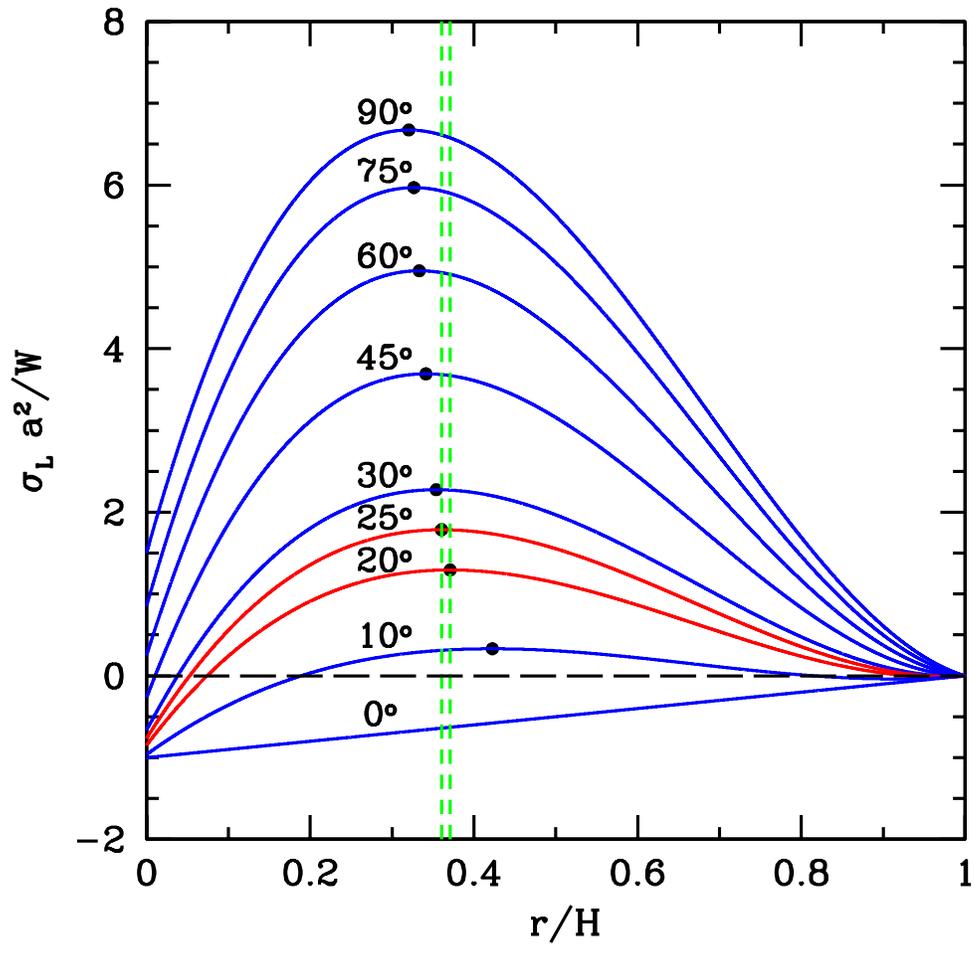

Fig. 3. The normalized longitudinal stress at the leading edge is shown as a function of the height fraction and for several angles. The maxima of the stress curves are marked by solid points. Curves in red are related to the breaking pattern of the structure shown in the previous figure.